\documentclass[fleqn,usenatbib]{mnras}

\usepackage{newtxtext,newtxmath}

\usepackage[T1]{fontenc}
\usepackage{ae,aecompl}

\usepackage{graphicx}	
\usepackage{amsmath}	
\usepackage{gensymb}
\usepackage{ulem}
\usepackage{color} 
\usepackage{subfiles} 
\usepackage[british]{babel}
\usepackage[utf8]{inputenc}

\definecolor{comment}{RGB}{200,0,0}

\newcommand{\widevec}[1]{\overrightarrow{#1}}

\title[Galaxy Zoo Builder: Morphological Dependence of Spiral galaxy Pitch Angle]{Galaxy Zoo Builder: Morphological Dependence of Spiral Galaxy Pitch Angle}

\author[T. Lingard et al.]{
  Timothy Lingard$^{1}$\thanks{E-mail: tklingard@gmail.com}, 
  Karen L. Masters$^{2}$, 
  Coleman Krawczyk$^{1}$, 
  Chris Lintott$^{3}$, 
  \newauthor
  Sandor Kruk$^{4}$, 
  Brooke Simmons$^{5}$, 
  William Keel$^{6}$, 
  Robert C. Nichol,$^{1}$ 
  \newauthor
  Elisabeth Baeten$^{7}$
  \\
  $^{1}$Institute of Cosmology and Gravitation, University of Portsmouth, Dennis Sciama Building, Burnaby Road, Portsmouth, PO1 3FX, UK\\
  $^{2}$Haverford College, 370 Lancaster Ave., Haverford, PA 19041, USA\\
  $^{3}$Oxford Astrophysics, Denys Wilkinson Building, Keble Road, Oxford, OX1 3RH, UK\\
  $^{4}$European Space Agency, ESTEC, Keplerlaan 1, NL-2201 AZ, Noordwijk, The Netherlands\\
  $^{5}$Physics Department, Lancaster University, Lancaster, LA1 4YB, UK\\
  $^{6}$Department of Physics \& Astronomy, University of Alabama, Tuscaloosa, AL 35457-0324, USAxs\\
  $^{7}$Independent Zooniverse Volunteer\\
}

\date{Accepted XXX. Received YYY; in original form ZZZ}

\pubyear{2021}

\hypersetup{draft}

\begin{document}
\label{firstpage}
\pagerange{\pageref{firstpage}--\pageref{lastpage}}
\maketitle

\begin{abstract}
  Spiral structure is ubiquitous in the Universe, and the pitch angle of arms in spiral galaxies provide an important observable in efforts to discriminate between different mechanisms of spiral arm formation and evolution. In this paper, we present a hierarchical Bayesian approach to galaxy pitch angle determination, using spiral arm data obtained through the \textit{Galaxy Builder} citizen science project. We present a new approach to deal with the large variations in pitch angle between different arms in a single galaxy, which obtains full posterior distributions on parameters. We make use of our pitch angles to examine previously reported links between bulge and bar strength and pitch angle, finding no correlation in our data  (with a caveat that we use observational proxies for both bulge size and bar strength which differ from other work). We test a recent model for spiral arm winding, which predicts uniformity of the cotangent of pitch angle between some unknown upper and lower limits, finding our observations are consistent with this model of transient and recurrent spiral pitch angle as long as the pitch angle at which most winding spirals dissipate or disappear is larger than 10$^\circ$.
\end{abstract}

\begin{keywords}
galaxies: evolution -- galaxies: spiral -- galaxies: photometry
\end{keywords}




\section{Introduction}
\begin{figure*}
  \includegraphics[width=15cm]{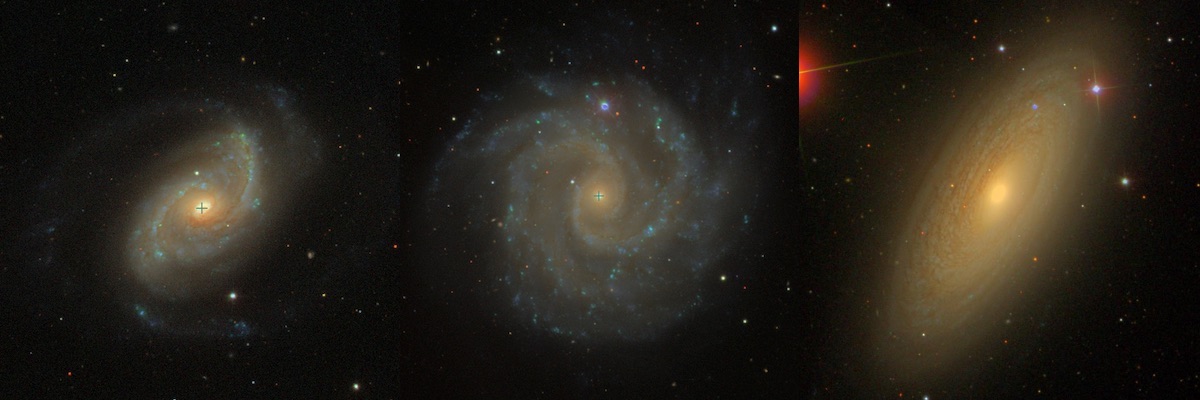}
  \caption{Examples of the different types of spiral galaxy present in the sky. The left column shows the grand design spiral NGC 5248. the middle shows the many-armed spiral NGC-3184 and the right shows the flocculent spiral NGC 2841. Images were taken with the Sloan Digital Sky Survey Telescope.}
  \label{fig:spiral-galaxy-types}
\end{figure*}

Spiral structure is present in a majority of massive galaxies (e.g. \citealt{1989gadv.book..151B}, \citealt{2008MNRAS.389.1179L}) yet the formation mechanisms through which spiral structure originates are still hotly debated (e.g. \citealt{2014PASA...31...35D}). Spirals are as diverse as the theories proposed to govern their evolution, from the quintessential pair of well-defined arcs of the grand design spiral, to the fragmented arm segments of the flocculent spiral, to the disjointed multi-armed spiral.
(\citealt{2011ApJ...737...32E}; examples of each type are shown in Figure \ref{fig:spiral-galaxy-types}). The Hubble classification scheme \citep{1926ApJ....64..321H} and its revisions and expansions \citep{1961hag..book.....S,1991rc3..book.....D} contain detailed variations of different types of spiral galaxy, divided by the presence of a bar and ordered by the openness of the spiral arms, the degree of resolution into condensations and the prominence of a central bulge. Building on this, \citet{1982MNRAS.201.1021E} found that flocculent spirals are more prevalent in unbarred, isolated galaxies. The presence of a bar, a binary companion or group membership result in a higher fraction of observed grand design spiral patterns.

Whatever kind of spiral is present in a disc galaxy, there is plenty of evidence to suggest that they have a significant role on the overall evolution of that galaxy. For example, a majority of the population of young stars in a galaxy are located in its spiral arms \citep{2011EAS....51...19E}, and there is evidence that spiral arms may trigger star formation \citep{2013A&A...560A..59C} perhaps via their ability to promote the growth of Giant Molecular Clouds \citep{2014IAUS..298..221D}. The rearrangement of disc gas and stars driven by spiral arms (e.g. \citealt{2018MNRAS.476.1561D}) may lead to the formation of disc-like bulges (commonly called ``pseudobulges''; e.g. \citealt{2004ARA&A..42..603K}), which are prevalent in most spiral galaxies, including those without bars \citep{2010ApJ...716..942F}. Studies of spiral morphology have also found interesting correlations with other galactic properties, such as a correlation between the tightness of spiral arms and central mass concentration (\citealt{2019ApJ...871..194Y}, though neither \citealt{2017MNRAS.472.2263H} nor \citealt{2019MNRAS.487.1808M} found such a relation in large samples). Spiral tightness is also observed to correlate with with rotation curve shape (\citealt{2005MNRAS.359.1065S}), with galaxies with rising rotation curves having more open spiral structure. These predictions and observations provide compelling reasons for continued investigation of the underlying rules and dynamics of spiral structure, as doing so is essential for understanding the secular evolution of disc galaxies.

Our current understanding of the mechanisms which drive spiral growth and evolution suggests that different forms of spiral arms in a galaxy may be triggered by different processes. Grand design spirals are thought to have undergone a tidal interaction \citep{2010MNRAS.403..625D,2017ApJ...834....7S}, be driven by a bar (as seen in gas simulations, \citealt{1976ApJ...209...53S,2008A&A...489..115R}, and suggested for stars by Manifold theory, \citealt{2006A&A...453...39R,2009MNRAS.394...67A,2009MNRAS.400.1706A}), or be obeying (quasi-stationary) density wave theory (QSDW theory), in which spiral arms are slowly evolving, ever-present structures in the disc (as first proposed by \citealt{1964ApJ...140..646L}). Flocculent spirals are thought to be formed through swing amplification (shearing of small gravitational instabilities in the disc), and be transient and recurrent in nature \citep{1966ApJ...146..810J}.

One of the fundamental assumptions of early work on spiral formation mechanisms (primarily QSDW) was that the disc of a galaxy, if unstable to spiral perturbations, would create a stable, static wave which would exist unchanged for many rotational periods \citep{1964ApJ...140..646L}. The motivation for static waves with small numbers of arms was primarily observational: most disc galaxies observed at the time showed spiral structure with low spiral arm numbers, suggesting that spirals exist for a long time or are continually rebuilt. This, in combination with theoretical arguments about the ``winding problem", motivated the original static density waves of \citet{1964ApJ...140..646L}, to which swing amplification was added by \citet{Toomre1981}  to provide a way to counteract the short lifetime of stellar density waves.

\begin{figure}
  \includegraphics[width=8.4cm]{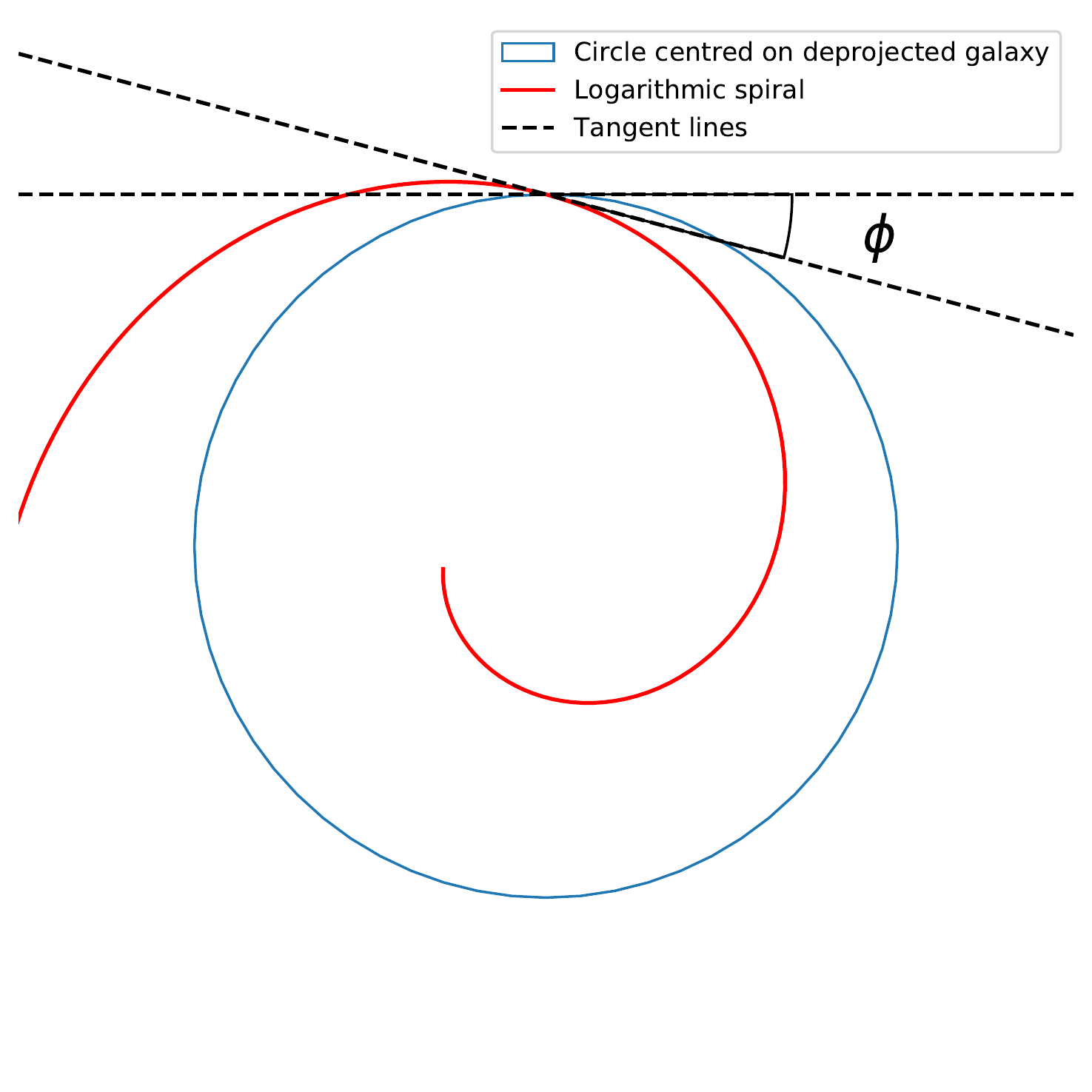}
  \caption{Illustration of the definition of pitch angle. It is given as $\phi = \tan^{-1}\left(\frac{\mathrm{d}r}{\mathrm{d}\theta}\ /\ r\right)$, or the angle between the spiral (red) and the tangent to a circle centred on the galaxy (blue).}
  \label{fig:pitch-angle-example}
\end{figure}

More recently, simulations demonstrate that spirals do not maintain a constant tightness (often quantified by pitch angle, the angle between the spiral and the tangent to a circle centred on the galaxy, \citealt{1987gady.book.....B}, illustrated in Figure \ref{fig:pitch-angle-example}), and instead wind-up over time due to the differential rotation of the disc \citep{2013ApJ...763...46B}. Recent research suggests that spirals arms are transient, and continually dissipate and re-form \citep{2014PASA...31...35D}. These spirals can be maintained through the same mechanisms that drive QSDW spirals (i.e. ``wave amplification by stimulated emission of radiation'', \citealt{1976ApJ...205..363M}; swing amplification, \citealt{1965MNRAS.130..125G}), but do not require the idealistic disc conditions required for the formation and maintenance of a stationary wave. The pitch angles of these transient spiral arms will decrease due to the differential rotation of the disc, with the density of the arm peaking at some critical pitch angle, before dissipating to be reformed.

In this dynamic picture of spiral arms, pitch angle monotonically decreases from a spiral arm's formation to its dissipation. As a particular example of this, \citet{2019MNRAS.490.1470P} propose a simple test of the winding of spiral arms, predicting that the cotangent of the pitch angle of a spiral arm ($\cot \phi$) evolves linearly with time. They found that the distribution of pitch angles of their sample of 86 galaxies was consistent with this prediction, which they present as evidence against QSDW theory in favour of the dynamic spirals produced in many simulations.

We aim to test this idea of spiral winding using data from the \textit{Galaxy Builder} citizen science project for the spiral galaxies present in \citet{2020arXiv200610450L}. We make use of Bayesian hierarchical modelling to measure galaxy pitch angle from the spiral arms produced by \textit{Galaxy Builder}. This methodology allows us to quantify the differences in pitch angles between arms in a single galaxy, as well as investigate the distribution of pitch angles in the galaxy population and investigate relationships between pitch angle and galaxy morphology.

Using Galaxy Zoo 2 data \citep{2013MNRAS.435.2835W} we further separate the galaxies by the presence and strength of a stellar bar. This will allow us to test simulations of gas in barred galaxies, which often demonstrate that bars can drive long-term spiral evolution \citep{2008A&A...489..115R}, or boost transient spiral structure \citep{2012MNRAS.426..167G}. Manifold theory \citep{2006A&A...453...39R,2009MNRAS.394...67A,2009MNRAS.400.1706A} is one attempt to determine the orbits of stars in bar-driven spiral arms: it proposes that stars in the vicinity of the unstable Lagrangian points at either end of the bar tend to escape along predictable orbits, governed by invariant manifolds. One of the primary factors influencing the shape of this invariant manifold is the relative strength of the non-axisymmetric forcing caused by the bar, with stronger bars resulting in spirals with larger pitch angles.

Many other galactic components may correlate with spiral morphology, including bulge fraction (\citealt{1975A&A....44..363Y}, \citealt{2013MNRAS.436.1074S}, \citealt{2019MNRAS.487.1808M}) and black hole mass (\citealt{2008ApJ...678L..93S}, \citealt{2017MNRAS.471.2187D}, \citealt{2019MS&E..571a2118A}). Larger bulges and more massive central black holes have both been observed to correlate with more tightly wound spiral arms. We can also test this with the data presented in this paper.

This paper is structured as follows: In Section \ref{section:method} we introduce methods to measure galaxy pitch angle, and present our sample, and our Bayesian hierarchical modelling method making use of \textit{Galaxy Builder} to estimate galaxy and population pitch angles. Section \ref{section:constraints-on-galaxy-phi} presents our general constraints on pitch angles in our section, while Section \ref{section:morphology-comparison} examines the correlation between pitch angle and bulge size implied by the Hubble sequence, and pitch angle and bar strength implied by Manifold theory. Section \ref{section:spiral_winding} investigates spiral arm winding using the test derived by \cite{2019MNRAS.490.1470P} (uniformity of galaxy pitch angle in $\cot\;\phi$). We provide a summary and conclusions in Section \ref{section:summary}.  Where necessary, we make use of $H_0 = 70\ \text{km}\ \text{s}^{-1}\ \text{Mpc}^{-1}$.

\section{Method}
\label{section:method}

\subsection{Measuring galaxy pitch angle}
Many methodologies have been proposed and implemented to measure spiral arm properties, including visual inspection (\citealt{2015A&A...582A..86H}), Fourier analysis (i.e. \textsc{2DFFT}, \citealt{2012ApJS..199...33D}), texture analysis (i.e. SpArcFiRe, \citealt{2014ApJ...790...87D}), and combinations of automated methods and human classifiers (\citealt{2017MNRAS.472.2263H}, \citealt{2020MNRAS.493.3854H}). One potentially underused method of obtaining measurements of spirals is through photometric fitting of spiral structure, as possible using tools such as \textsc{GALFIT} \citep{2010AJ....139.2097P} and \textit{Galaxy Builder} \citep{2020arXiv200610450L}. These methods attempt to separate light from an image of a galaxy into distinct subcomponents, such as a galaxy disc, bulge, bar and spiral arms, generally finding the optimum solution using computational optimisation. This optimisation process, however, is often not robust for complex, many-component models and requires significant supervision to converge to a physically meaningful result \citep{Gao2017:1709.00746v1}. \citet{2020arXiv200610450L} proposed a solution to this problem through the use of citizen science to provide priors on parameters used in computational fitting.

A common assumption when measuring galaxy pitch angle is that observed spiral arms have a constant pitch angle with radius (e.g. \citealt{2012ApJS..199...33D,2013MNRAS.436.1074S,2014ApJ...790...87D}). Spirals of this kind are known as logarithmic spirals and are described by

\begin{equation}
  \label{eq:log-spiral}
r = A\,e^{\theta\tan\phi},
\end{equation}
where $\phi$ is the arm's pitch angle, $A$ is an amplitude coefficient and $\theta$ is the polar coordinate. Different arms in a galaxy could have different values of $\phi$, however for each arm, $\phi$ is assumed to be constant with radius. One method used to obtain a pitch angle of a galaxy is to fit logarithmic spirals to individually identified arm segments and take the weighted mean of their pitch angles (which often vary by upwards of $10^\circ$, \citealt{2014ApJ...790...87D}). Weighting is determined by the length of the arc segment, with longer arms being assigned higher weights, i.e. for a galaxy where we have identified $N$ arm segments, each with length $L_i$ and pitch angle $\phi_i$

\begin{equation}
  \phi_\mathrm{gal} = \left(\sum_{i=1}^{N}L_i\right)^{-1}\sum_{i=1}^{N}L_i \phi_i.
\end{equation}

The most commonly used measurement of uncertainty of length-weighted pitch angles is the unweighted sample variance between the arm segments which were identified.

A notable drawback of length-weighted pitch angle is sensitivity to the number and quality of the spiral arm segments; \citet{2017MNRAS.472.2263H} found that only 15\% of the arm segments which were identified using \textsc{SpArcFiRe} \citep{2014ApJ...790...87D} were identified as ``good'' matches to real spiral arms by citizen science classifiers.

Fourier analysis in one- and two-dimensions (as performed by \citealt{2019A&A...631A..94D}, \citealt{2012ApJS..199...33D}, \citealt{2018MNRAS.474.2594M}, and dating back to the seminal work of \citealt{ConsidereAthanassoula1988}) is another widely used method of computationally obtaining galaxy pitch angles. Two-dimensional Fourier methods generally decompose a deprojected image of a galaxy into a superposition of logarithmic spirals between inner and outer annuli \citep{2012ApJS..199...33D} and report the pitch angle with the highest amplitude as the galaxy's pitch angle. \citet{2020MNRAS.493.3854H} combined Fourier analysis of spiral galaxies with a visual tracing of spiral arms, successfully eliminating observed bias in a sample of toy images of galaxies. It is unclear how the variation between pitch angles of individual arms impacts this measurement. We note that while this method is able to model non-logarithmic spirals -- as a sum of logarithmic spirals with differing pitch angles, most applications use models which assume that the pitch angle is constant with radius, in some cases picking regions of a galaxy in which this is true - e.g. see Section 4.3.2 of \citealt{2012ApJS..199...33D}.

\subsection{The galaxy sample}
\begin{figure*}
  \includegraphics[width=17.7cm]{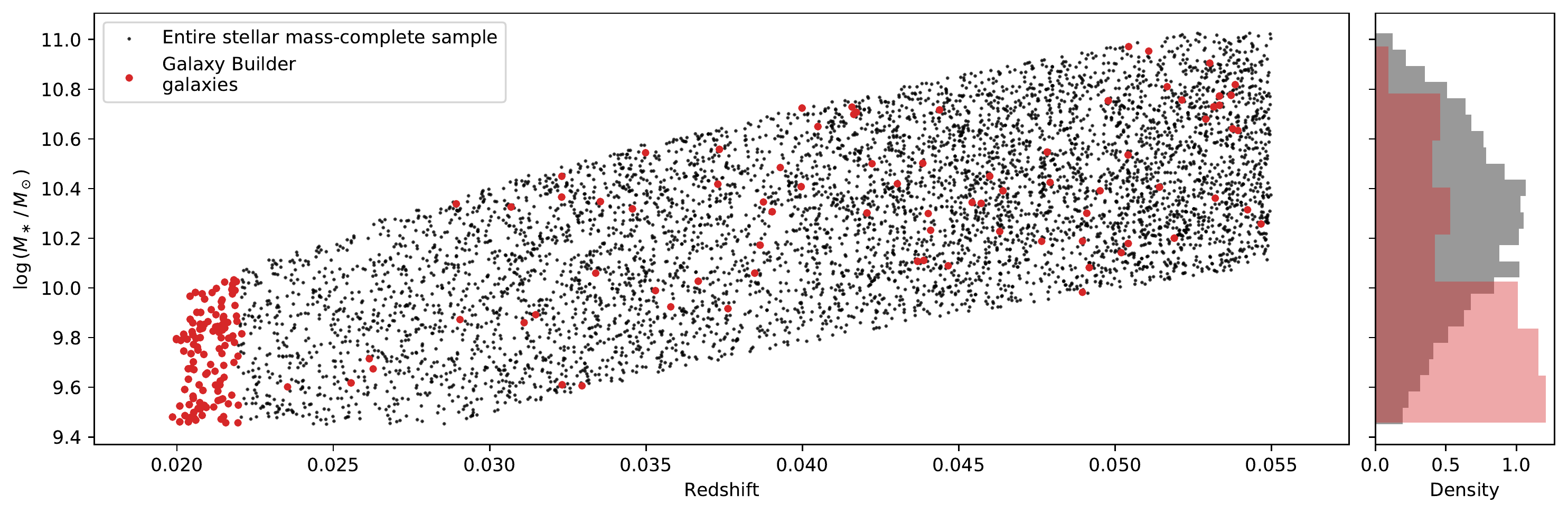}
  \caption{A plot of redshift against stellar mass for the \textit{stellar mass-complete sample} from \citet{2017MNRAS.472.2263H}; the subset we use for analysis in Galaxy Builder are shown in red. At right we show a histogram of the stellar masses. This Figure is identical to Figure 4 from \citet{2020arXiv200610450L} who use the same sample.}
  \label{fig:stellarmass}
\end{figure*}

The galaxies analysed in this paper are those for which photometric models were obtained in \citet{2020arXiv200610450L}. These are a subset of the \textit{stellar mass-complete sample} in \citet{2017MNRAS.472.2263H}, a sample of low-redshift ($0.02 < z < 0.055$) face-on spiral galaxies selected using data from the NASA-Sloan Atlas \citep{2011AJ....142...31B} and Galaxy Zoo 2 \citep{2013MNRAS.435.2835W}. The \textit{stellar mass-complete sample} ranged in stellar mass from $9.45 < \log(M_* / M_\odot) < 11.05$, with most of the sample between $9.5 < \log(M_* / M_\odot) < 10.0$. A histogram of stellar masses for our subset can be see in Section \ref{section:constraints-on-galaxy-phi}, where variation with stellar mass, to check the impact of this limited mass range, is also investigated. For the reader's convenience we also reproduce Figure 4 from \citet{2020arXiv200610450L} here (see Figure \ref{fig:stellarmass}) which shows the redshift and stellar mass distribution of our analysis sample compared to the full \textit{stellar mass-complete sample} from \citet{2017MNRAS.472.2263H}. Our choice (see \citet{2020arXiv200610450L} for details) to prefer lower redshift galaxies for {\it Galaxy Builder} analysis is clear in the mass distribution which results in a sample favouring galaxies $9.5 < \log(M_* / M_\odot) < 10.0$, and includes a smaller number of spirals with masses up to  $\log(M_* / M_\odot) = 11.05$.

Some galaxies in \citet{2020arXiv200610450L} were shown to volunteers a second time in a repeat validation subset to create a second aggregate model used to test internal consistency. Section 3.2 of \citet{2020arXiv200610450L} presents a comparison of these classifications to investigate volunteer consistency. We can see that arm number is highly reliable within $\pm1$ (93\% galaxies have arm number counts which agree in this range; 53\% have exactly identical arm number counts). In this work, we combine the 30 classifications of galaxies in this validation subset with the 30 original classifications. Clustering of drawn spiral arms and cleaning of points was then performed as detailed in \citet{2020arXiv200610450L}. We remove any galaxies for which no spiral arms were identified, resulting in a hierarchical data structure of 129 galaxies, 247 spiral arms and 238,433 points. This breaks down further to 68 galaxies (53\% of the sample) having two arms identified (meaning that they were marked by enough users to cluster into an arm), 19 (15\%) with three such arms, 4 (3\%) with four, and the remainder (38 or 29\%) with a single identified arm. We believe that our reported number of arms per galaxy is in many cases an underestimate, and since the most common number of arms is two, this most often results in a single arm being measured when two are present. The most common way we miss an arm is if the clustering did not converge when users have appeared to identify a second arm.
For example, we find by visual inspection that just three galaxies in the sample are truly one-armed spirals, while in the remaining 35 galaxies for which only one arm was identified following clustering, the other arms were simply not recovered due to noise in the dataset. Given this, we do not recommend using these statistics to make general conclusions about the number of arms per galaxy. However, as our hierarchical model incorporates the uncertainty involved with missing spiral arms, the results related to pitch angle should not be significantly affected.

Spiral arm points are deprojected to a face-on orientation using the disc inclination and position angle obtained through photometric model fitting performed in \citet{2020arXiv200610450L}. Arms are individually corrected to all have the same chirality (a pitch angle greater than or equal to zero) using the logarithmic spiral fit in \citet{2020arXiv200610450L}. This was achieved by multiplying the polar coordinate $\theta$ by $-1$ for arms identified as winding counter-clockwise.

\subsection{Bayesian modelling of spiral arms in \textit{Galaxy Builder}}
\label{section:bhsm-model}
In this section, we lay out our Bayesian hierarchical model for galaxy pitch angle. We fit directly to clustered, cleaned points from polylines drawn in \textit{Galaxy Builder}, deprojected and unwrapped to polar coordinates. We fit a logarithmic spiral to each clustered arm (examples are shown in Figure \ref{fig:example-spiral-fits}), with the pitch angles of multiple arms in a single galaxy being drawn from a single parent distribution.

Logarithmic spirals have the desirable properties of a constant pitch angle and a small number of free parameters. For this first analysis of the \textit{Galaxy Builder} models we choose to make use of it here without an explicit comparison to other models. A simple visual inspection of the fitted logarithmic spirals suggests that it is an appropriate model, however, a comparison of a logarithmic spiral profile to other spiral forms (i.e. Archimedean or polynomial) is another important piece of work, outside of the scope of this research, as it has been reported that galaxy arms do not have constant pitch angles (\citealt{1981AJ.....86.1847K}; \citealt{2009MNRAS.397..164R}).

As suggested by spiral formation models which correlate a galaxy wide pitch angle with galaxy wide properties, we will assume that a given galaxy has some preferred value for arm pitch angle, $\phi_\mathrm{gal}$, and that the pitch angles of spiral arms in that galaxy, $\phi_\mathrm{arm}$, are constant with radius (giving logarithmic spirals) and drawn from a normal distribution centred on $\phi_\mathrm{gal}$, with some spread $\sigma_\mathrm{gal}$ common to all galaxies. We truncate the normal distribution of spiral arm pitch angles in a single galaxy between the physical limits of {0\degree} (a ring) and {90\degree} (a ``spoke''), giving

\begin{equation}
\phi_\mathrm{arm} \sim \mathrm{TruncatedNormal}(\phi_\mathrm{gal}, \sigma_\mathrm{gal}, \mathrm{min}=0, \mathrm{max}=90).
\end{equation}

The choice to assume all galaxies show the same inter-arm variation in pitch angle (represented by a common value of $\sigma_\mathrm{gal}$ across all galaxies) was motivated by our small sample size and the low number of arms measured per galaxy. With this sample size we do not find, nor expect to be sensitive to variations in this parameter. It is possible that it does vary between galaxies, and that this variation is physically interesting. Several authors have previously made attempts to measure this parameter.  In a seminal work, \citet{1981AJ.....86.1847K} fit logarithmic spirals to 113 nearby (NGC) galaxies, and note the dominant error in average galaxy pitch angle comes from inter-arm variation, which they measure to have an average value of 5$^\circ$. \citet{2014ApJ...790...87D} is primarily a machine learning method paper, and while details on the galaxies investigated are not clear, their Table 1 presents the median difference in pitch angle between pairs of arms with different lengths, which varies from 14.5$^\circ$ in very short arms, to 2.6$^\circ$ in the longest traced arms. It is unclear how much this encodes error in their method versus real variation in the galaxy population. Within our own Milky Way, \citet{Vallee2015} did a meta analysis and comparison of several technique to conclude a range of 12-14$^\circ$ (i.e. 2$^\circ$) was reasonable for all Milky Way spiral arms. In a very detailed study of four very nearby spirals, \citet{HonigRead2015} conclude there are large variations of pitch angles  between spirals, and among arms in a given spiral, but made no comments as to if the variation was consistent with being constant. Further investigation of this issue in a larger \textit{Galaxy Builder} sample would be an interesting follow-up project.

We assume that the observed points in a \textit{Galaxy Builder} spiral arm, once deprojected, follow a logarithmic spiral with gaussian radial error $\sigma_r$,

\begin{equation}
\widetilde{r_\mathrm{arm}} = \exp\left(\widevec{\theta_\mathrm{arm}}\tan\phi_\mathrm{arm} + c_\mathrm{arm}\right).
\end{equation}

Where $\widetilde{r_\mathrm{arm}}$ is the model's prediction for the radii of the deprojected points in a \textit{Galaxy Builder} arm ($\widevec{r_\mathrm{arm}}$), $c_\mathrm{arm}$ is the amplitude parameter (equivalent to $A$ in Equation \ref{eq:log-spiral}), and $\widevec{\theta_\mathrm{arm}}$ is the polar angles of the points.

We choose hyperpriors over $\phi_\mathrm{gal}$, $\sigma_\mathrm{gal}$, $c_\mathrm{arm}$ and $\sigma_\mathrm{r}$ of

\begin{align}
  \phi_\mathrm{gal} &\sim \mathrm{Uniform}(\mathrm{min}=0, \mathrm{max}=90),\\
  \sigma_\mathrm{gal} &\sim \mathrm{InverseGamma}(\alpha=2,\,\beta=20),\\
  c_\mathrm{arm} &\sim \mathrm{Cauchy}(\alpha=0,\,\beta=10),\\
  \sigma_r &\sim \mathrm{InverseGamma}(\alpha=2, \beta=0.5).
\end{align}

These are conservative priors, which are not expected to have significant impact on the results. The inverse gamma distribution is used to aid the convergence of the Hamiltonian Monte Carlo (HMC) algorithm used (discussed later). The Cauchy distribution is equivalent to the Student's t-distribution with one degree of freedom, and was chosen due to its fatter tails than the normal distribution. Our likelihood function for $N$ arms, each with $n_\mathrm{arm}$ points, is

\begin{equation}
  \mathcal{L} = \prod_{\mathrm{arm}=1}^{N}\left(2\pi\sigma_r^2\right)^{-n_\mathrm{arm}/2}
  \exp\left(-\frac{||\widevec{r_\mathrm{arm}}\,-\,\widetilde{r_\mathrm{arm}}||^2}{2\sigma_r^2}\right).
\end{equation}

We assume that the radial error is Gaussian for simplicity of analysis, however, Shapiro-Wilk tests on the residuals of the logarithmic spirals fit in \citet{2020arXiv200610450L} suggest that this is not a good assumption, and a more robust likelihood (such as the Student's t-distribution) would possibly more appropriate.

To perform inference, we make use of the No-U-Turn-Sampler (NUTS, \citealt{2011arXiv1111.4246H}), implemented in PYMC3\footnote{\url{https://docs.pymc.io/}}, an open-source probabilistic programming framework written in Python \citep{pymc3_paper}. To aid the convergence of MC chains, we scale the radii of deprojected points to have unit variance.

\section{Results}
\subsection{Constraints on galaxy pitch angle}
\label{section:constraints-on-galaxy-phi}
\begin{figure*}
  \includegraphics[width=17.7cm]{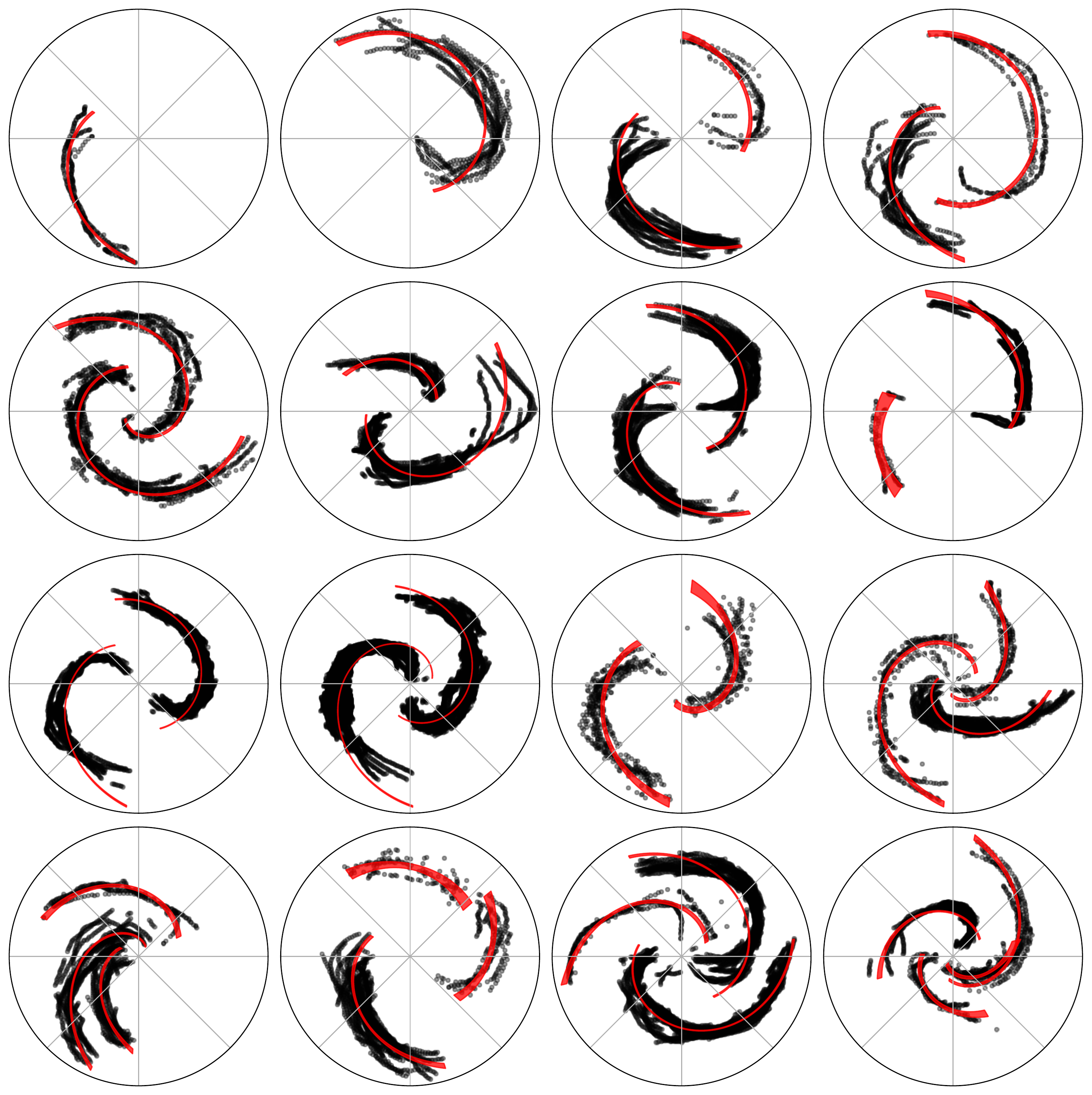}
  \caption{Examples of spiral profiles fit using the hierarchical model described in Section \ref{section:bhsm-model}. Deprojected points from \textit{Galaxy Builder} clustered, cleaned spiral arms are shown in black; fit logarithmic spiral arms are shown in red, with the width of the line corresponding to the $2\sigma$ interval on predicted values of $\widetilde{r_\mathrm{arm}}$. The two one-armed spirals in the top left panels are instances where the spiral clustering algorithm failed to identify all spiral arms present in the galaxy.}
  \label{fig:example-spiral-fits}
\end{figure*}

\begin{figure*}
  \includegraphics[width=17.7cm]{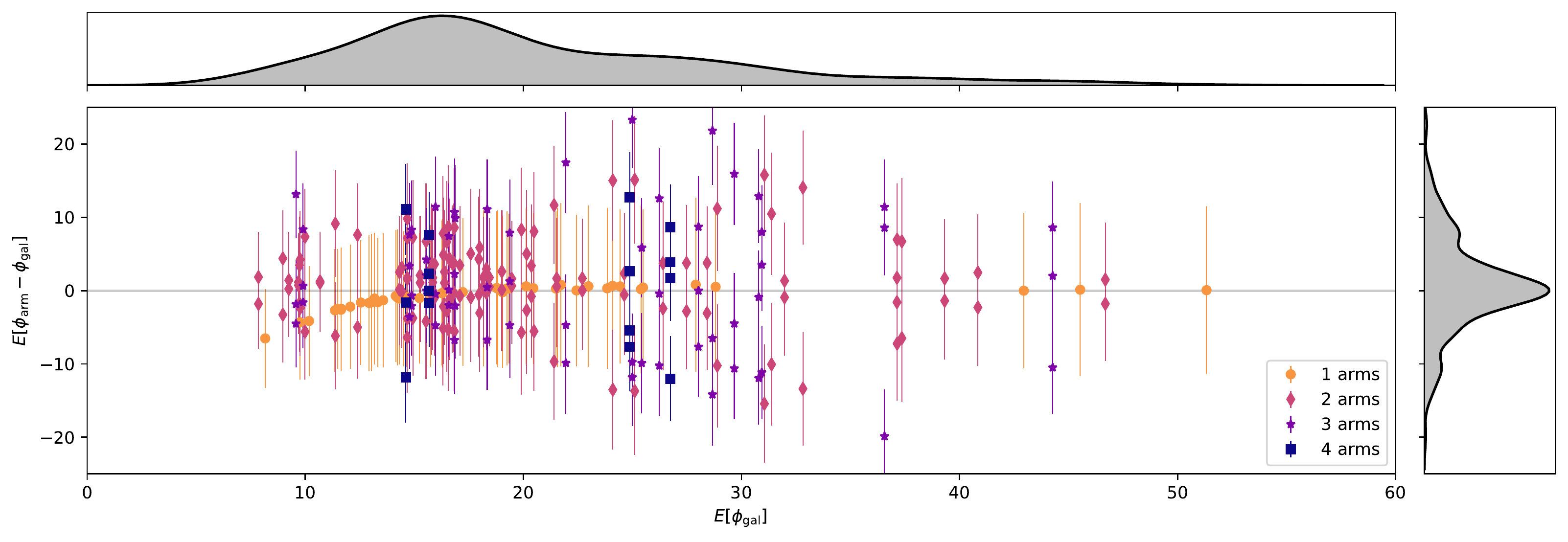}
  \caption{Scatter plot showing how arm pitch angle compares to galaxy pitch angle for galaxies with different pitch angles and number of arms. The top panel shows a Gaussian KDE for $E[\phi_\mathrm{gal}]$, and the right panel shows a Gaussian KDE for $E[\phi_\mathrm{arm} - \phi_\mathrm{gal}]$. The galaxy pitch angle is consistent with the mean of its arms, with large scatter and a slight bias against values near the lower bound of $0$ due to the lower limit applied.}
  \label{fig:arm-pa-spread}
\end{figure*}

Our hierarchical model identifies the pitch angle of individual arms ($\phi_\mathrm{arm}$) with posterior standard deviations less than {1.6\degree} for 95\% of arms, assuming no error on disc inclination and position angle. This is illustrated well by the small uncertainties on fit spiral arms in Figure \ref{fig:example-spiral-fits}. The pitch angle of a galaxy as a whole ($\phi_\mathrm{gal}$), however, is not well constrained. This is primarily a result of only having pitch angles measurements for a small number of arms per galaxy, and reflects the difficulty in providing a single value for the pitch angle of a galaxy containing individual arms with very different pitch angles. For galaxies with two arms identified in \textit{Galaxy Builder}, we have a mean uncertainty of ($\sigma_{\phi_\mathrm{gal}}$) of  {7.9\degree}, which decreases to {6.8\degree} and {6.0\degree} for galaxies with three and four arms respectively. This is roughly consistent with the standard error on the mean for a galaxy with $N$ arms,

\begin{equation}
  \sigma_{\phi_\mathrm{gal}} = \frac{\sigma_\mathrm{gal}}{\sqrt{N}},
\end{equation}
where $\sigma_\mathrm{gal}$ is our measure of inter-arm variability of pitch angle and has a posterior distribution of $11.0^\circ\pm 0.9^\circ$. This inter-arm variability is similar to that found by \citet{1981AJ.....86.1847K} and \citet{2014ApJ...790...87D} and emphasises the need for fitting algorithms to not assume all arms have the same pitch angle. The spread of arm pitch angle from the mean galaxy pitch angle can be seen in Figure \ref{fig:arm-pa-spread}, with points colour-coded by the number of arms measured for a galaxy. We see a slight drop in the expectation values of galaxy pitch angle ($E[\phi_\mathrm{gal}]$) compared to the expectation of arm pitch angles ($E[\phi_\mathrm{arm}]$) at small galaxy pitch angles, which is caused by a combination of the truncation of $\phi_\mathrm{gal}$ at {0\degree} and the large spread (so the mean value differs from the mode as the distribution is highly skewed).

\begin{figure}
  \includegraphics[width=9cm]{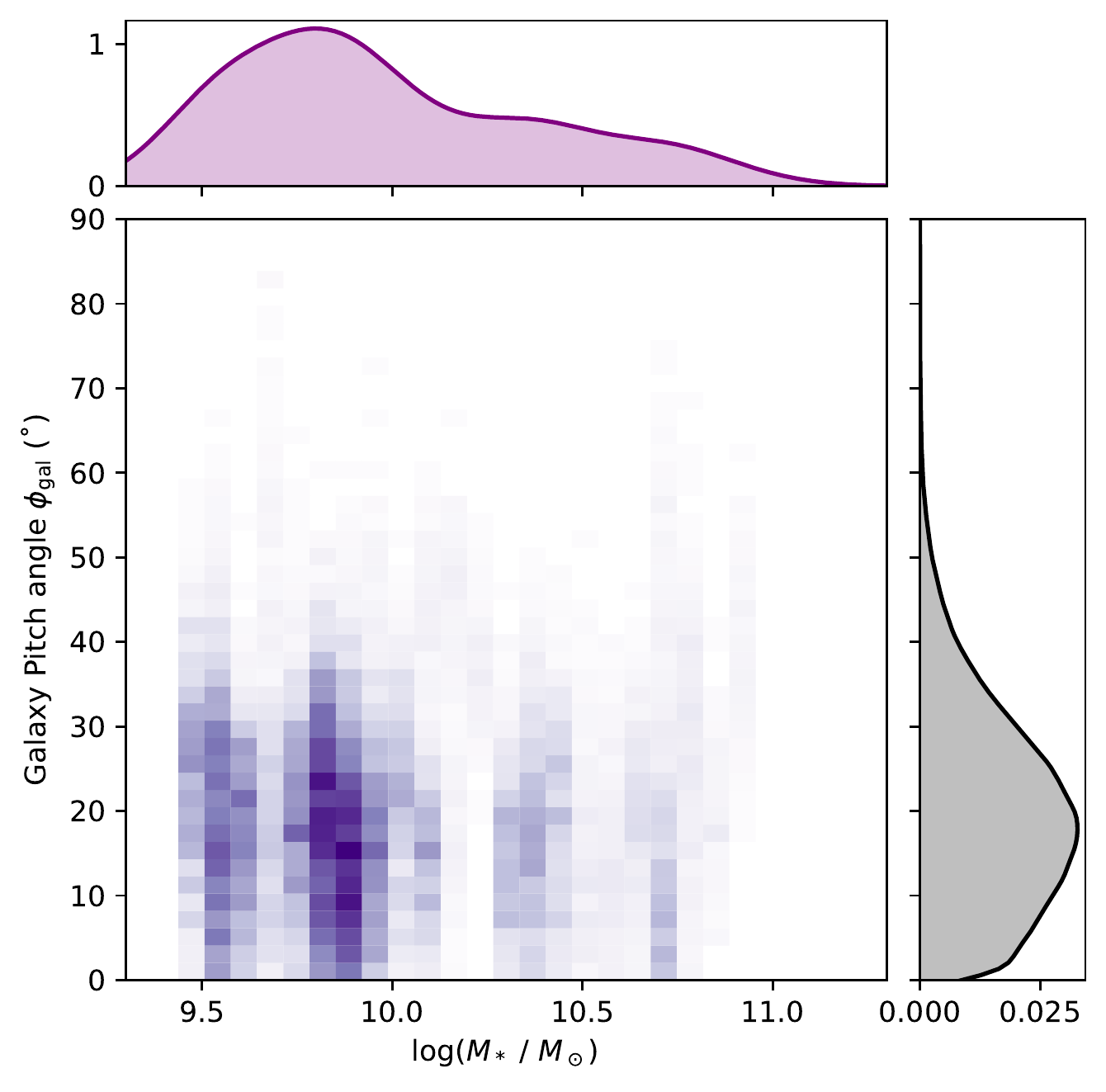}
  \caption{The stellar mass distribution of the sample and the galaxy average pitch angle distribution shown as a 2D histogram (centre) and also projected along each axis. }
  \label{fig:stellar-mass-phigal}
\end{figure}

In Figure \ref{fig:stellar-mass-phigal} we present the stellar mass distribution of the sample, and investigate how the global galaxy pitch angle depends on this parameter. The majority of our sample has stellar masses $9.5 < \log(M_* / M_\odot) < 10.0$, and galaxy average pitch angles 10$^\circ$--20$^\circ$. This should be remembered in our physical interpretation of other results. We observe no significant trend of the global pitch angle with mass, although there is a hint that more massive spirals may on average have more tightly wound arms.

\subsection{Dependence of pitch angle on galaxy morphology}
\label{section:morphology-comparison}

To test the possible progenitor distribution of our estimated arm pitch angles, we repeatedly perform an Anderson-Darling test (\citealt{10.2307/2286009}, implemented in \textsc{Scipy}, \citealt{scipy-paper}) over each draw present in the MC trace, resulting in a distribution of Anderson-Darling statistics. We will refer to this test as the \textit{marginalized Anderson-Darling test}. We also make use of the two-sample Anderson-Darling \citep{doi:10.1080/01621459.1987.10478517} test in a similar manner.

We make use of Galaxy Zoo 2 data for morphological comparison. Two of the galaxies in our sample could not be matched to Galaxy Zoo 2 data, and as such have been dropped from this comparison (leaving 127 galaxies).

\subsubsection{Pitch angle vs. Bulge size}
\label{section:morphology-comparison-bulge}

Morphological classification commonly links bulge size to spiral tightness, and such a link is implied by the Hubble Sequence (\citealt{2005ARA&A..43..581S}, \citealt{2009MNRAS.393.1531G}, \citealt{2013seg..book..155B}), although small bulge Sa galaxies have been noted for decades (e.g. for a review see \citet{2005ARA&A..43..581S}; this is also noted in \citealt{2019MNRAS.487.1808M}). Some recent studies have indeed reported a link between measured spiral galaxy pitch angle and bulge size (i.e. \citealt{2019ApJ...873...85D}), while others have not found any significant correlation \citep{2019MNRAS.487.1808M}. The differing results of this may depend on the details of how both the bulge size, and spiral pitch angles are measured and suggest further investigation is needed. We investigate this relationship here using a measure of bulge prominence from Galaxy Zoo 2, as Equation 3 in \citet{2019MNRAS.487.1808M}:

\begin{equation}
  B_\mathrm{avg} = 0.2\times p_\mathrm{just\ noticeable} + 0.8\times p_\mathrm{obvious} + 1.0\times p_\mathrm{dominant},
\end{equation}
where $p_\mathrm{just\ noticeable}$, $p_\mathrm{obvious}$ and $p_\mathrm{dominant}$ are the fractions of classifications indicating the galaxy's bulge was ``just noticeable'', ``obvious'' or ``dominant'' respectively.

We see no correlation between galaxy pitch angle derived from the hierarchical model and $B_\mathrm{avg}$. The Pearson correlation coefficient between the expectation value of galaxy pitch angle ($E[\phi_\mathrm{gal}]$) and $B_\mathrm{avg}$ is 0.00 (with a p-value of 0.95).

\begin{figure*}
  \includegraphics[width=17.7cm]{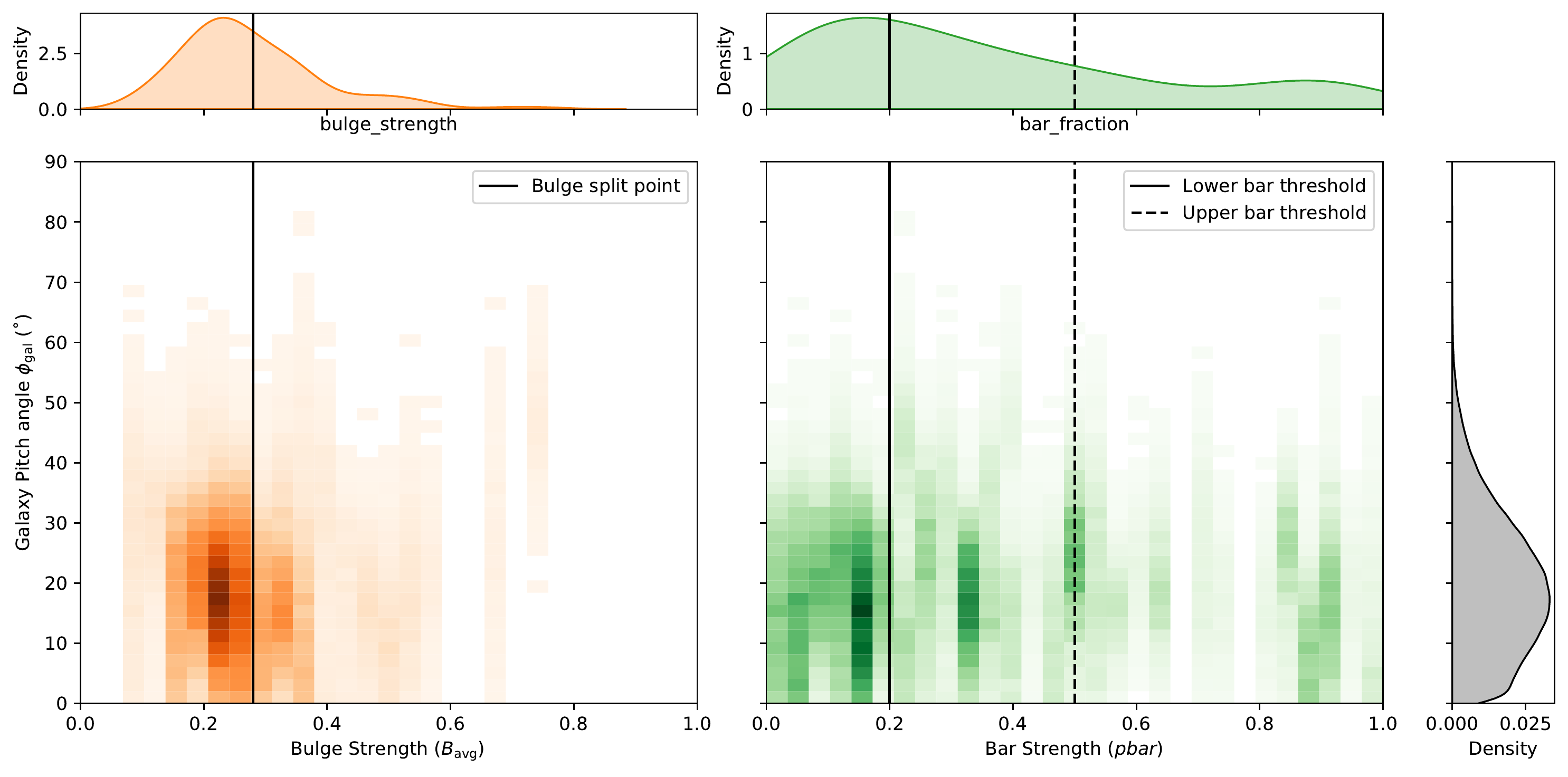}
  \caption{Density plot showing bulge strength ($B_\mathrm{avg}$; left, orange) and bar strength ($p_\mathrm{bar}$; right, green) against galaxy pitch angle ($\phi_\mathrm{gal}$). Split points for the marginalized Anderson-Darling tests are labelled. There is no statistically significant relationship for either bulge or bar strength.}
  \label{fig:bulge-bar-pa-hist}
\end{figure*}

We separate our sample into galaxies with weaker bulges ($B_\mathrm{avg} < 0.28$, 79 galaxies) and those with stronger bulges ($B_\mathrm{avg} \ge 0.28$, 48 galaxies), to test whether their pitch angles could be drawn from significantly different distributions. A marginalized two-sample Anderson-Darling test comparing the distributions of $\phi_\mathrm{gal}$ for the samples does not find evidence that galaxy pitch angles were drawn from different distributions: we reject the null hypothesis at the 1\% level for only 1\% of the samples. Similarly comparing arm pitch angles for galaxies in the different samples results in not rejecting the null hypothesis at the 1\% level for any of the samples. The distributions of the Anderson-Darling test statistic for $\phi_\mathrm{gal}$ and $\phi_\mathrm{arm}$ are shown in the upper panel of Figure \ref{fig:ad-morphology-test} in blue and orange respectively.

One limitation of this result is that our sample does not contain many galaxies with dominant bulges: $B_\mathrm{avg}$ only varied from 0.09 to 0.75 (the allowed maximum being 1.0), with only four galaxies having $B_\mathrm{avg} > 0.5$. The split point of 0.28 was also chosen to produce evenly sized comparison samples rather than from some physical motivation. However, the lack of any form of correlation implies that there is no evidence in our data for the link between bulge size and pitch angle predicted by the Hubble sequence and observed in other studies.

\subsubsection{Pitch angle vs. Bar Strength}
\label{section:morphology-comparison-bar}

One of the predictions of Manifold theory is that pitch angle increases with bar strength as evaluated by the Quadrupole moment, $Q$ at the Lagrangian $L_1$ point (a value that differs from typical ``bar strength'', which is this value averaged over all radii in the bar; \citealt{2009MNRAS.400.1706A}). In this work we do not have any similar measurement of bar strength and we note that \citet{2009MNRAS.400.1706A} caution that other measures of bar strength may not show this relation; we also do not have significant numbers of strongly barred galaxies in our sample. However in an attempt to investigate this relationship in our data, we make use of Galaxy Zoo 2's bar fraction ($p_\mathrm{bar}$), which has been demonstrated to be a good measure of bar length \citep{2013MNRAS.435.2835W} and bar strength \citep{2012MNRAS.423.1485S,2012MNRAS.424.2180M,2018MNRAS.473.4731K} and therefore a good measure of the torque applied on the disc gas.

\begin{figure*}
  \includegraphics[width=17.7cm]{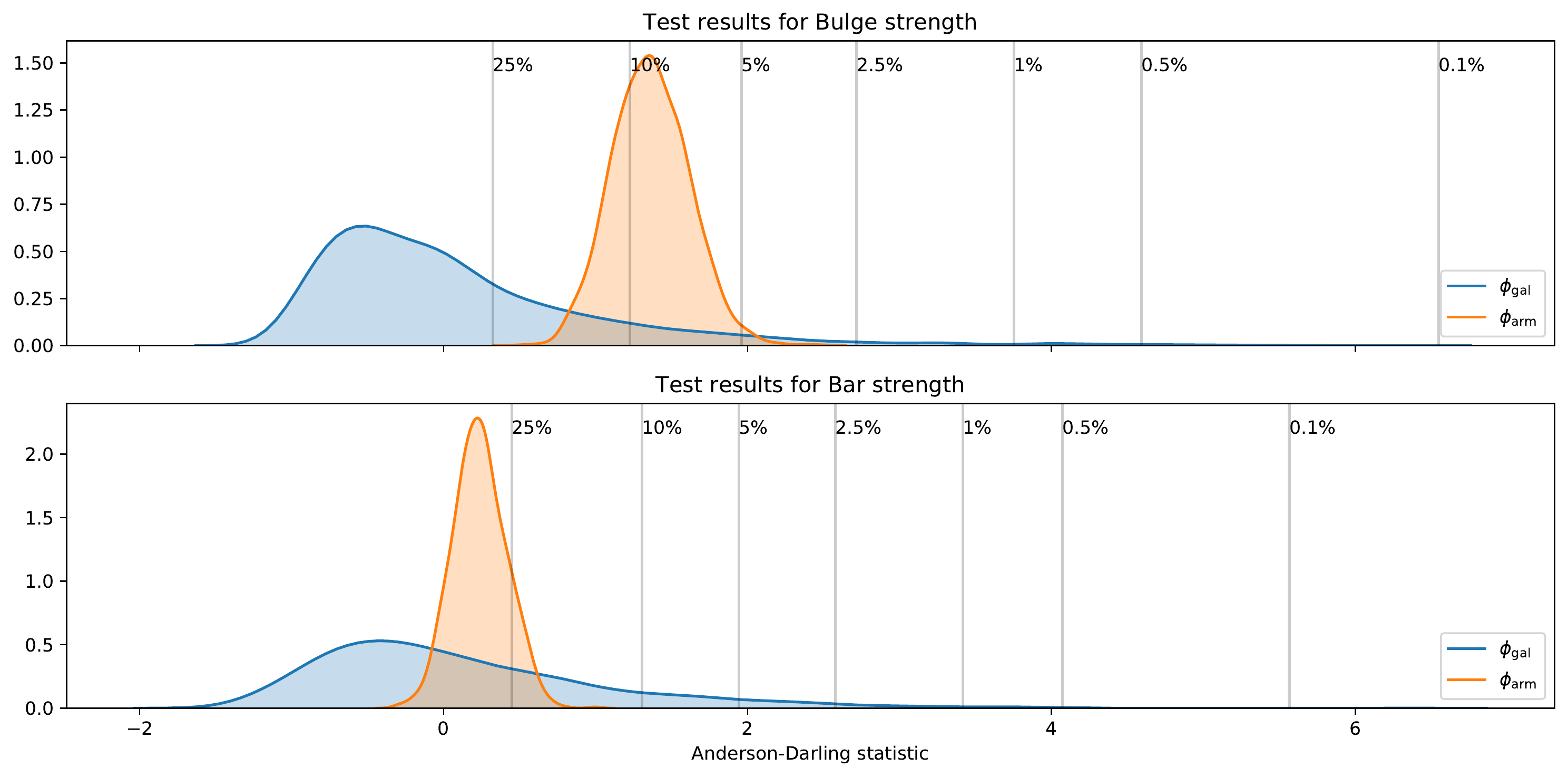}
  \caption{The results of marginalized two-sample Anderson-Darling tests examining whether pitch angles ($\phi_\mathrm{gal}$ in blue and $\phi_\mathrm{gal}$ in orange) for galaxies with $B_\mathrm{avg} < 0.28$ and $B_\mathrm{avg} \ge 0.28$ are drawn from the same distribution (top panel), and the results of marginalized three-sample Anderson-Darling tests for galaxies with no bar ($p_\mathrm{bar} < 0.2$), a weak bar ($0.2 \le p_\mathrm{bar} \le 0.5$) and a strong bar ($p_\mathrm{bar} > 0.5$) (bottom panel). Confidence intervals are shown, with moving rightwards indicating more confidence in rejecting the null hypothesis that the compared values were drawn from the same parent distribution. We cannot reject the null hypothesis at the 1\% level for any of the tests conducted, meaning there is no evidence in this sample that bulge size or bar strength impacts pitch angle.}
  \label{fig:ad-morphology-test}
\end{figure*}

We do not observe a correlation between $p_\mathrm{bar}$ and $E[\phi_\mathrm{gal}]$ (Pearson correlation coefficient of -0.05, with a p-value of 0.54). Following \citet{2012MNRAS.424.2180M} and \citet{2012MNRAS.423.1485S}, we separate the sample into galaxies without a bar ($p_\mathrm{bar} < 0.2$, 50 galaxies), with a weak bar ($0.2 \le p_\mathrm{bar} \le 0.5$, 44 galaxies) and with a strong bar ($p_\mathrm{bar} > 0.5$, 33 galaxies). Performing marginalized three-sample Anderson-Darling tests does not find that pitch angles ($\phi_\mathrm{gal}$ or $\phi_\mathrm{arm}$) of galaxies with different bar strengths were drawn from different distributions; we do not reject the null hypothesis at the 1\% level for any samples for the test of $\phi_\mathrm{gal}$, and at the 10\% level for the test of $\phi_\mathrm{arm}$. The distributions of Anderson-Darling test statistic is shown in the lower panel of Figure \ref{fig:ad-morphology-test}.

The fact that we do not find any link between our measure of bar strength (based on the prominence of the bar in Galaxy Zoo 2) and pitch angle is suggestive that there is actually no link between bar strength and pitch angle, which would exclude Manifold theory as the primary mechanism driving the evolution of the spirals in our sample. However since we use only a proxy for bar strength, which has not been well tested, this is not conclusive.

\subsection{Spiral Winding}
\label{section:spiral_winding}

For transient and recurrent spiral arms driven by self-gravity, \citet{2019MNRAS.490.1470P} suggest that spiral patterns form at some maximum pitch angle ($\phi_\mathrm{max}$), continually wind up over time and finally dissipate at some minimum pitch angle ($\phi_\mathrm{min}$). They propose that, under a set of very simple assumptions, the evolution of pitch angle would be governed by

\begin{equation}
  \label{eq:winding}
  \cot{\phi} = \left[R\frac{\mathrm{d}\Omega_p}{\mathrm{d}R}\right](t - t_0) + \cot{\phi_\mathrm{max}},
\end{equation}
where $\Omega_p$ is the radially dependant pattern speed of the spiral arm and $t_0$ is the initial time at which it formed.

In QSDW theory, the pattern speed $\Omega_p$ is a constant in R, as spiral arms obey rigid-body rotation. If $\Omega_p$ instead varies with radius we would expect $\cot{\phi}$ to be uniformly distributed between $\cot{\phi_\mathrm{max}}$ and $\cot{\phi_\mathrm{min}}$.  The model presented in \citet{2019MNRAS.490.1470P} does not give any physical justification for what $\cot{\phi_\mathrm{max}}$ and $\cot{\phi_\mathrm{min}}$ should be.

To test this theory, \citet{2019MNRAS.490.1470P} used a Kolmogorov-Smirnov test to examine whether a sample of 113 galaxies with measured pitch angles was likely to have been drawn from a distribution uniform in its cotangent. Pitch angles were measured by \citet{2019ApJ...871..194Y} using discrete Fourier transformations in one- and two-dimensions, and as such do not account for inter-arm variations. They conclude the model works within limits of $\cot{\phi} \in [1.00, 4.75]$ (roughly $11.9^\circ < \phi < 45.0^\circ$), motivated by examination of the data.

\begin{figure*}
  \includegraphics[width=17.7cm]{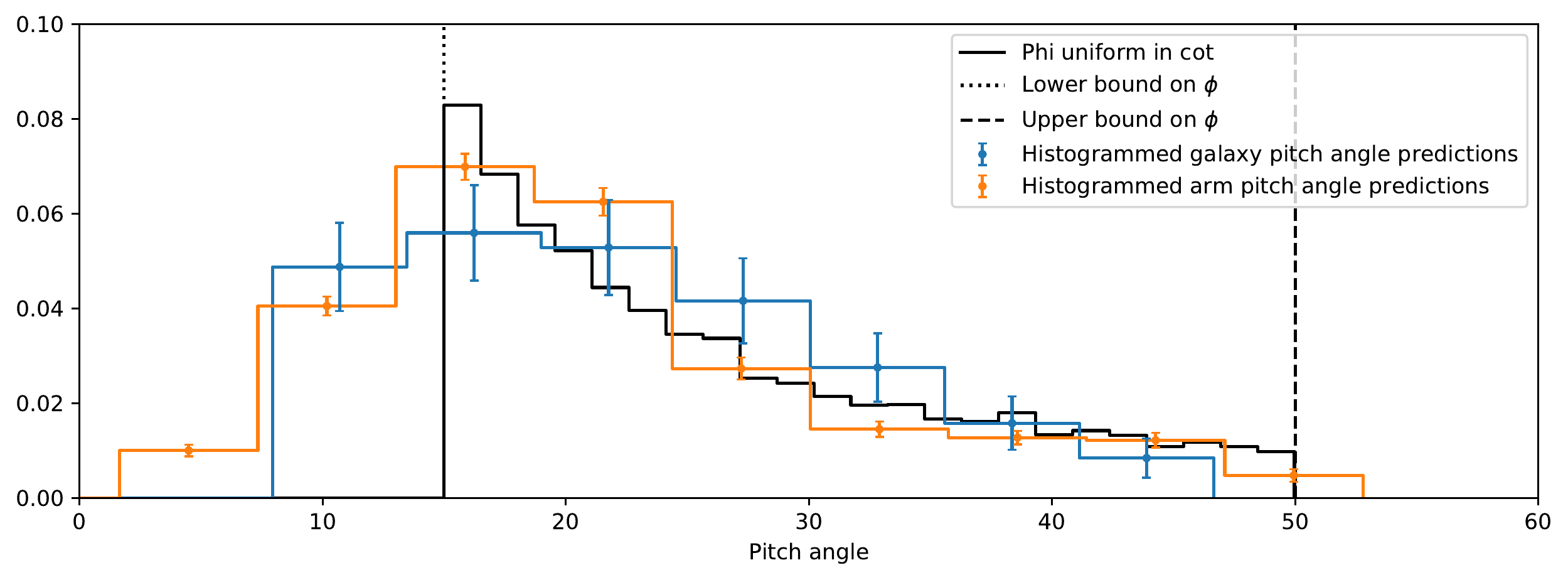}
  \caption{The distributions of pitch angles (orange and blue) relative to one uniform in $\cot\phi$ (black). Histograms have been normalised by the area between the limits such that they are comparable. The histogram was recalculated with identical bins for each posterior sample of $\phi_\mathrm{gal}$ and $\phi_\mathrm{arm}$, we plot the mean value of each bin, with the sample standard deviation shown as error bars. It is evident that the distributions are very similar between the chosen limits.}
  \label{fig:pa-cot-distributions}
\end{figure*}

We perform a similar test in this work, using our sample and methods. We will make use of the marginalized Anderson-Darling test described above, and examine winding on a per-arm basis, as well as a per-galaxy basis. Observation of the distribution of arm pitch angles in our sample (Figure \ref{fig:pa-cot-distributions}) suggests they are close to uniform in cotangent within limits of $15^\circ < \phi < 50.0^\circ$.

\subsubsection{Galaxy pitch angle}

Testing the uniformity of $\cot\phi_\mathrm{gal}$ between {15\degree} and {50\degree} using a marginalized Anderson-Darling test results in rejecting the null hypothesis at the 1\% level for just 5\% of samples, with a large spread in observed test values. The full distribution of Anderson-Darling statistics can be seen in Figure \ref{fig:ad-cot-test}. The large spread in results is caused by the large uncertainties in $\phi_\mathrm{gal}$.

This result suggests that our data are consistent with a cot-uniform source distribution for galaxy pitch angle, but the large uncertainty in $\phi_\mathrm{gal}$ makes it difficult to make any conclusive statements. This result is also highly sensitive to the lower limit of $\phi$: decreasing it to {10\degree} results in us rejecting the cot-uniform model at greater than the 0.1\% level for 96\% of the posterior samples. We can conclude from this that the \citet{2019MNRAS.490.1470P} cot-uniform model is an adequate fit to the data, as long as the minimum pitch angle, $\cot{\phi_\mathrm{min}}$, at which the majority of winding dissipate or disappear is  $\phi_\mathrm{min} > $ {10\degree}, and more confidently $\phi_\mathrm{min} = $ {15\degree}. We reiterate that there is no prediction in \citet{2019MNRAS.490.1470P} as to what this minimum pitch angle should be, so our observation constrains the allowed range.

\subsubsection{Arm Pitch angle}
The inconclusive result for $\phi_\mathrm{gal}$ is perhaps unsurprising: were we to assume that spiral arms are transient and recurrent instabilities, there is little reason for all of the arms to be at precisely the same evolutionary stage at the same time. This is supported by the large observed spread in inter-arm pitch angles (Section \ref{section:constraints-on-galaxy-phi}).

If we assume instead that spirals form and wind independently inside a galaxy, and that their evolution over time can be described by Equation \ref{eq:winding}, the distribution of the cotangent of pitch angles of individual arms should be uniform between our limits, rather than that of the galaxy's pitch angle as a whole.

\begin{figure*}
  \includegraphics[width=17.7cm]{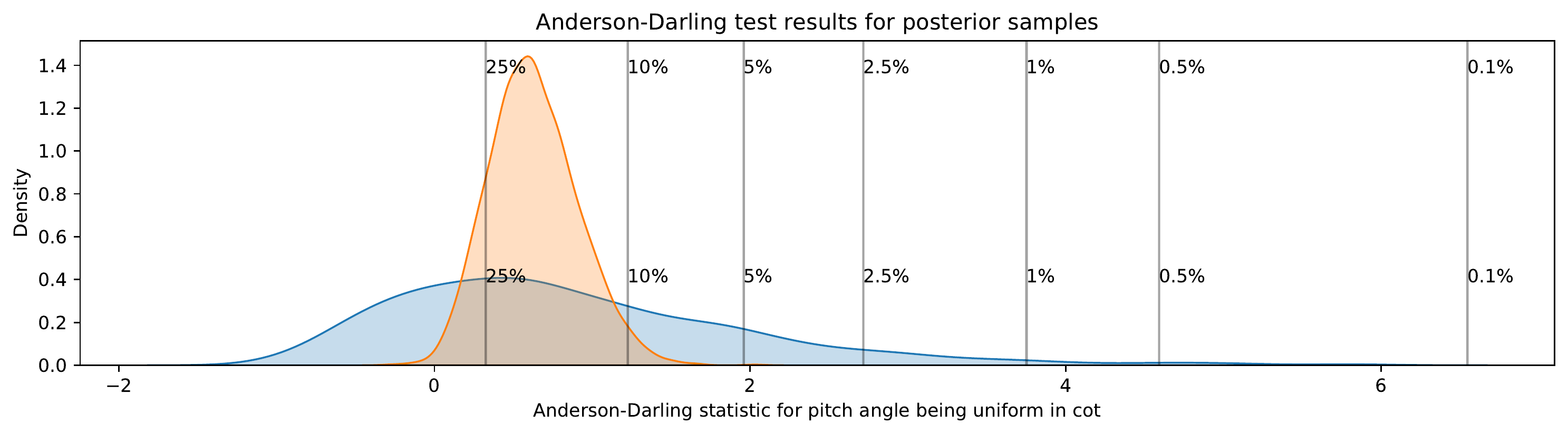}
  \caption{The results of a marginalized Anderson-Darling test for uniformity in $\cot$ for $\phi_\mathrm{gal}$ (blue) and $\phi_\mathrm{arm}$ (orange), with values corresponding to various confidence intervals shown. Moving rightwards on the x-axis implies greater confidence in rejecting the null hypothesis that the sample was drawn from a distribution uniform in $\cot$ between $15^\circ < \phi < 50.0^\circ$. In this instance, we would not be able to reject the null hypothesis at the 1\% level for either $\phi_\mathrm{gal}$ or $\phi_\mathrm{arm}$, meaning our sample is consistent with a cot uniform distribution. The larger error in $\phi_\mathrm{gal}$  means that this result is more significant for $\phi_\mathrm{arm}$, which is also physically motivated, as arms can wind independently.}
  \label{fig:ad-cot-test}
\end{figure*}

Using the marginalized Anderson-Darling test we cannot reject the null hypothesis at even the 5\% level for any of the possible realizations of arm pitch angle. The resulting distribution of Anderson-Darling statistics is shown in in the lower panel of Figure \ref{fig:ad-cot-test}. This result is highly consistent with the model for spiral winding proposed by \citet{2019MNRAS.490.1470P} with $\cot{\phi_\mathrm{min}} = $ {15\degree} and can be interpreted as evidence that spirals are formed through local disc perturbations, and are primarily governed by local forces.

\section{Summary and Conclusions}
\label{section:summary}
This paper presents a new Bayesian approach to estimate galaxy pitch angle, making use of citizen science results to measure spiral arms through photometric modelling. We introduce an adaptation of the Anderson-Darling test, which we name the \textit{marginalized Anderson-Darling test}, to incorporate full Bayesian posterior probabilities and use this test to investigate theories governing spiral formation and evolution.

The hierarchical Bayesian approach implemented in this paper allows a more thorough examination of pitch angle than length-weighted pitch angle calculation obtaining posterior distributions of measured parameters. It better accounts for the large variations observed in inter-arm pitch angle than Fourier analysis, which assumes all arms in a given symmetric mode have the same pitch angle. In this work, we find that the mean inter-arm difference in pitch angle is $11.0^\circ\pm 0.9^\circ$.

There is no evidence in our data for the link between bulge size and pitch angle predicted by the Hubble sequence and observed in other studies (see Section \ref{section:morphology-comparison-bulge}).

We do not find any link between our measure of bar strength and pitch angle in our sample. However, rather that a direct measure of bar strength, we make use of an available parameter which correlates with bar strength, so at best this observation is suggestive that the primary mechanism driving the evolution of the spirals in our sample may not be Manifold theory (see Section \ref{section:morphology-comparison-bar}). Since this is not the measure of bar strength predicted to correlate with pitch angle by \citet{2009MNRAS.400.1706A}, and those authors caution that the details of the bar strength measure can wash out the predicted correlation, this is not strong evidence against Manifold theory models.

Our results are consistent with spiral winding of the form described by \citet{2019MNRAS.490.1470P}, in which spiral arms are transient and recurrent, evolve through mechanisms such as swing-amplification \citep{1965MNRAS.130..125G} and which wind up over time. This model predict a distribution of pitch angles that is uniform in cotangent space across some range. No prediction is provided as to what that range should be. Our data are consistent with this model, if the minimum pitch angle is $\phi_\mathrm{min} = $ {15\degree}, but rule it out if the minimum pitch angle is $\phi_\mathrm{min} = $ {10\degree}. The assumptions of this model of spiral winding are highly simplistic, and it leaves many unanswered questions: what determines the limits on $\phi$? Is the spiral arm equally apparent at all pitch angles, or is a selection effect present? Our observations suggest that any further development of this model needs to predict that the minimum pitch angle, $\phi_\mathrm{min}>${10\degree}.This result is also not evidence against QSDW, as our distribution of pitch angles may be dictated by other factors such as disc shear.

In this work, we assume that spiral arms are equally likely to be identified and recovered at all pitch angles, which suggests the absence of galaxies at low pitch angles is not due to an inability of us to measure such arms. This is not an unfair assumption given the amount of human effort that went into obtaining spiral arm measurements (more so than any other pitch angle measurement method, with each galaxy receiving at least 30 human classifications). The galaxy sample used is a random subset of a volume limited sample (see Figure \ref{fig:stellarmass}), and is comparable in size to those used in other similar studies (\citealt{2013MNRAS.436.1074S}, \citealt{2019ApJ...871..194Y}, \citealt{2019MNRAS.490.1470P}). The sample covers a range of masses ($9.45 < \log(M_* / M_\odot) < 11.05$) and spiral types, however it is possible that tightly wound spirals are preferentially missed by the pre-selection from \textit{Galaxy Zoo}, if they are less obviously identified as having spirals at these distances ($0.02<z<0.055$). The \citet{1981AJ.....86.1847K} sample of 113 much more nearby spirals (all at $z<0.019$ and most at $z<0.009$), includes several with arms at much lower pitch angles, but very few arms which are loosely wound (none with $\cot \phi > 31^\circ$), meaning it does not match the same $\cot \phi$ constant model well, although as at the time the sample was set by data availability, it is unclear how conclusive this is, and \citet{1981AJ.....86.1847K} note the incompleteness of their sample for open armed spirals. In a future version of {\it Galaxy Builder} we intend to include this sample as a comparison set.

We have presented evidence that the methodology proposed here is a robust solution to the problems facing investigation of spiral morphology, namely that of reliably identifying spiral arms, and properly accounting for the spread in pitch angles of arms within a galaxy. This is one of the largest samples for which this test has been done and is scaleable to larger samples; such a sample would make possible further comparisons, such as splitting galaxies into spiral type (grand design / many-armed / flocculent), examining the differences between populations, investigating if the interarm spread depends on other galaxy properties.

The processes governing the formation and evolution of spiral arms are complicated, but the prevalence of spiral galaxies in the Universe, their impact for understanding star formation, and the spiral nature of our own Milky Way, makes investigating their dynamics of fundamental importance to the scientific aims of understanding, predicting and explaining the nature of the cosmos.


\section{Acknowledgements}
This publication made use of SDSS-I/II data. Funding for the SDSS and SDSS-II was provided by the Alfred P. Sloan Foundation, the Participating Institutions, the National Science Foundation, the U.S. Department of Energy, the National Aeronautics and Space Administration, the Japanese Monbukagakusho, the Max Planck Society, and the Higher Education Funding Council for England. The SDSS Web Site is \url{http://www.sdss.org/}.

This publication uses data generated via the Zooniverse.org platform, development of which is funded by generous support, including a Global Impact Award from Google, and by a grant from the Alfred P. Sloan Foundation. We would also like to thank the 2,340 volunteers who have submitted classifications to the \textit{Galaxy Builder} project, especially user EliabethB, whose presence on the \textit{Galaxy Builder} forum on top of a large number of galaxies modelled has been a huge help.

This project was partially funded by a Google Faculty Research Award to Karen Masters (\url{https://ai.google/research/outreach/faculty-research-awards/}), and Timothy Lingard acknowledges studentship funding from the Science and Technology Facilities Council (ST/N504245/1).

\section{Data Availability}
The data and software analysis underlying this paper are available on GitHub at \url{https://github.com/tingard/gzbuilder_results} and \url{https://github.com/tingard/hierarchical-modelling-of-spiral-pitch-angle}.



\bibliographystyle{mnras}
\bibliography{bibliography} 




\bsp	
\label{lastpage}
\end{document}